\documentclass[
    ,final            
  ]
  {aipproc}

\layoutstyle{6x9}

\usepackage{amssymb}
\usepackage{graphicx}

\newcommand{\bi}[1]{\bibitem{#1}}

\newcommand{\FF}{{\mathcal F}}

\newcommand{\bgi}{\begin{itemize}}
\newcommand{\eni}{\end{itemize}}

\newcommand{\bb}{\begin{thebibliography}}
\newcommand{\eb}{\end{thebibliography}}

\newcommand{\nn}{\nonumber \\}
\newcommand{\bwt}{\begin{widetext}}
\newcommand{\ewt}{\end{widetext}}
\newcommand{\be}{\begin{equation}}
\newcommand{\ee}{\end{equation}}
\newcommand{\ba}{\begin{eqnarray}}
\newcommand{\ea}{\end{eqnarray}}
\newcommand{\baa}{\begin{eqnarray*}}
\newcommand{\eaa}{\end{eqnarray*}}
\newcommand{\bea}{\begin{eqnarray}}
\newcommand{\eea}{\end{eqnarray}}

\newcommand{\bslash}{{\mathcal B} \hskip -0.6em /}

\newfont{\fib}{cmfi10 at 10pt}

\newcommand{\eg}{{\it e.g.}\ }

\begin{document}

\title{Novel Azimuthal Asymmetries in 
Drell Yan and Semi-inclusive Deep Inelastic Scattering}

\classification{ 13.87.Fh  13.60.-r  13.88.+e  14.20.Dh 
}
\keywords {$T$-odd effects, Transversity}

\author{Leonard P.  Gamberg}{
  address={Division of Science,
Penn State-Berks Lehigh Valley College, 
Reading, PA 19610, USA}
}

\author{Gary R. Goldstein}{
  address={Department of Physics and Astronomy, 
Tufts University, Medford, MA 02156, USA}
}

\author{Karo A. Oganessyan}{
  address={  INFN-Laboratori Nazionali di Frascati, E. Fermi 40, 
00044 Frascati, Italy }
}

\begin{abstract}
We consider the leading and sub-leading 
twist $T$-odd and even contributions to the 
$\cos 2\phi$ azimuthal asymmetry in unpolarized
dilepton production in Drell-Yan Scattering.  
We estimate the contributions' effects  at $500\ {\rm GeV}$,
$ 50 \, {\rm GeV}$, and $25\, {\rm GeV}$ energies
in the framework of the 
parton model using a quark diquark-spectator model of the nucleon to approximate
the soft contributions.

\end{abstract}

\maketitle


\section{Introduction}
One of the most  interesting results in spin physics has been
the discovery of a class of chirally odd quark distribution functions. Considerable attention has been focused on the transversity or 
covariant transverse spin distribution 
$h_1$ which provides information on the quark transverse 
spin distribution in a transversely polarized 
nucleon~\cite{RS,JJ91}.  Chiral odd distribution functions 
require a quark helicity flip and are
difficult to probe in inclusive deep inelastic 
scattering due to the helicity conserving property 
of  quantum chromodynamics (QCD) interactions.
However, when two hadrons participate in the scattering 
process, the nucleon's transversity can be accessed; for example,
 the double transverse spin asymmetry 
in Drell-Yan scattering~\cite{RS,RHIC}. 
Alternatively, transversity can be probed
in semi-inclusive deep inelastic scattering (SIDIS) where outgoing
hadrons are produced  in the current fragmentation region~\cite{cnpb93,HERMES}.
In this  case the cross sections and distribution functions are
sensitive to the transverse momentum of the quarks.
Sensitivity to transverse 
momentum dependence in parton distribution functions leads to a class of
leading twist spin dependent effects
which are  $T$-odd~\cite{sivers,ANSL,BM,gold_gamb}. 
The  distribution functions exist by virtue of non-zero parton 
transverse momenta and would 
vanish at tree-level in any $T$-conserving model of hadrons and 
quarks. In this sense they are similar to the decay amplitudes for 
hadrons that involve single spin asymmetries which are non-zero due to 
final (and/or initial) state strong interactions (FSI)~\cite{cnpb93}. 

Such a $T$-odd distribution
was proposed by Sivers in the context of Drell-Yan 
scattering~\cite{sivers}. How such a distribution could actually 
arise without violating conservation laws remained an open question 
for some years. Recently, however, a mechanism was introduced that 
could produce that Sivers effect without violating invariance principles. 
A reaction mechanism in terms of FSI (and using the quark-diquark model 
of the nucleon) was used to calculate this effect
by Brodsky, Hwang and Schmidt~\cite{BHS} (BHS). 
That calculation was subsequently
recast in terms of a color gauge invariant
treatment of transverse momentum dependent distribution 
functions~\cite{ji,cplb02,pij}.  Gluon loop corrections to tree-level 
calculations for the transversity of quarks and hadrons 
emerge from the gauge link insertion into 
the formal definitions of the transversity distributions. 
At first order in the strong coupling $\alpha_s$ the correction involves the 
single gluon approximation to FSI, a single loop diagram.

These FSI corrections, implemented in  parton model  
inspired calculations, give rise to novel single spin and azimuthal 
asymmetries in SIDIS~\cite{BHS,gold_gamb,gamb_gold_ogan1,gamb_gold_ogan2,dis03,
BMY,gambsa,BSY} 
and Drell-Yan processes~\cite{DB,BBH} resulting
in significant asymmetries.  This 
is all the more interesting in light of the fact that in
the perturbative QCD regime of quark and gluon dynamics such 
asymmetries are expected to be small; that is,
of the order, $\alpha_s m_q/\sqrt{s}$~\cite{kane,dharma}. One $T$-odd 
transversity distribution function (introduced by Boer and Mulders~\cite{BM}) 
is  $h_1^\perp$, which depends on variables $x, Q^2$ and $p_\perp$ 
of the struck 
quark; it measures the amount of quark transversity from an unpolarized 
nucleon. 
At moderate energies such $T$-odd effects arising from
the non-perturbative $h_1^\perp$
may be the source
of non-trivial $\cos2\phi$ asymmetries 
in SIDIS~\cite{e665,gamb_gold_ogan1,dis03}
and in Drell-Yan scattering~\cite{conway,DB,BBH}.

In this letter we will report our results on SIDIS~\cite{gamb_gold_ogan1,gamb_gold_ogan2,dis03} and present new
results  on the Drell-Yan process~\cite{prep}. 
The latter process is interesting 
in light of the possibility of accessing this transvesity property of quarks 
in the proposed proton anti-proton experiments at Darmstat 
GSI~\cite{pax},
where an anti-proton beam is an ideal tool for studying 
transversity due to the dominance of valence quark effects. 
We demonstrate  
at proposed energies that both the $T$-odd and sub-leading
twist $T$-even distributions in $\bar{p}\ p\rightarrow \ell\ \ell^+\ X$
provide contributions to the $\cos2\phi $ asymmetry that are significant.  
We estimate that
 the sub-leading twist contribution is a non-trivial fraction of the
large leading twist $T$-odd contribution using
the parton model. 
\section{The Azimuthal Asymmetries in Unpolarized SIDIS and 
Drell Yan Scattering}
The angular asymmetries that can arise in {\em unpolarized} Drell-Yan 
scattering (\eg $\bar{p}+p \to \mu^- \mu^+ +X$) and SIDIS
( $e+p \to e'hX$) are
obtained from the differential cross section expressions:
\ba
\frac{1}{\sigma}\frac{d\sigma}{d\Omega}=\frac{3}{4\pi}\frac{1}
{\lambda+3}\left( 1+\lambda\cos^2\theta + \mu \sin^2\theta \cos\phi 
+ \frac{\nu}{2} \sin^2\theta \cos 2\phi\right)
\label{d-y_sigma}
\ea
and 
\begin{equation}
\frac{d\sigma}{dxdydzdP^2_{h\perp}d\phi_h}=  A+B+C\cos\phi+D\cos2\phi,
\label{CS}
\end{equation}
respectively~\cite{CS,CAHN,LM}.  In the Drell-Yan process
 the angles refer to the lepton pair orientation in their rest 
frame relative to the boost direction and the initial hadron's 
plane, the Collins Soper frame~\cite{CS}. 
The dependence on the other independent variables,  
$s, x, m_{\mu\mu}^2, q_T$, is suppressed, but the asymmetry 
parameters, $\lambda, \mu, \nu$, depend on those variables. 
In SIDIS,   the azimuthal angle  refers to the 
relative angle between the hadron production plane and the
lepton scattering  plane. $A, \ B, \ C$, 
and $D$ are functions of $x, y, z, Q^2, 
|\mathbf{P}_{h\perp}|$. 
It is especially interesting that the $\cos 2\phi$ azimuthal asymmetry 
in Drell Yan depends on the $T$-odd distribution $h_1^\perp$ and its 
anti-quark distribution, $\bar{h}_1^\perp$;  whereas in SIDIS one 
essentially replaces the anti-quark
distribution with the Collins function, $H_1^\perp$~\cite{cnpb93}. 
\subsubsection{$\cos 2\phi$ Azimuthal Asymmetry in Drell Yan}
 To leading 
order in $Q^2$, {\em i.e.} leading twist, the asymmetry $\nu$ is given by~\cite{DB}
\be
\nu_2 = \frac{\sum_a e^2_a
{\FF} 
\left[ 
(2 \hat{\mathbf{h}}\cdot \mathbf{k}_{\perp }\cdot\hat{\mathbf{h}}\cdot 
\mathbf{p}_{\perp }
 - \mathbf{p}_\perp \cdot \mathbf{k}_\perp)
h_1^\perp(x, \mathbf{k}_\perp)\bar{h}_1^\perp(\bar{x}, \mathbf{p}_\perp)/
(M_1 M_2)
\right]
}
{\sum_a e_a^2 {\FF} f_1(x, \mathbf{k}_\perp) \bar{f}_1(\bar{x}, \mathbf{p}_\perp)}
\label{nu_eqn}
\ee
where the convolution is,
$
{\FF}\equiv \int d^2\mathbf{p}_\perp d^2\mathbf{k}_\perp 
\delta^2(\mathbf{p}_\perp + 
\mathbf{k}_\perp - \mathbf{q}_\perp) f^a(x,\mathbf{k}_\perp) 
\bar{f}^a(\bar{x},\mathbf{p}_\perp).
$
As pointed out by Collins and Soper~\cite{CS}, well before 
$h_1^{\perp}$ was identified, there 
is a higher twist $T$-even contribution to the $\cos2\phi$ asymmetry which
is not small at center of mass energies of $50\ {\rm GeV}^2$
\be
\nu_4=\frac{\frac{1}{Q^2}\sum_a e^2_a{\FF}
\left[\left(2\left(\hat{\mathbf{h}}\cdot\left(\mathbf{k}_\perp -\mathbf{p}_\perp \right)\right)^2
-\left(\mathbf{k}_\perp-\mathbf{p}_\perp\right)^2\right)
f_1(x, \mathbf{k}_\perp) \bar{f}_1(\bar{x}, \mathbf{p}_\perp)
\right]}
{\sum_a e_a^2 {\FF}\left[ f_1(x, \mathbf{k}_\perp) 
\bar{f}_1(\bar{x}, \mathbf{p}_\perp)\right]}.
\ee
\vspace{0.7cm}
\begin{figure}[h]
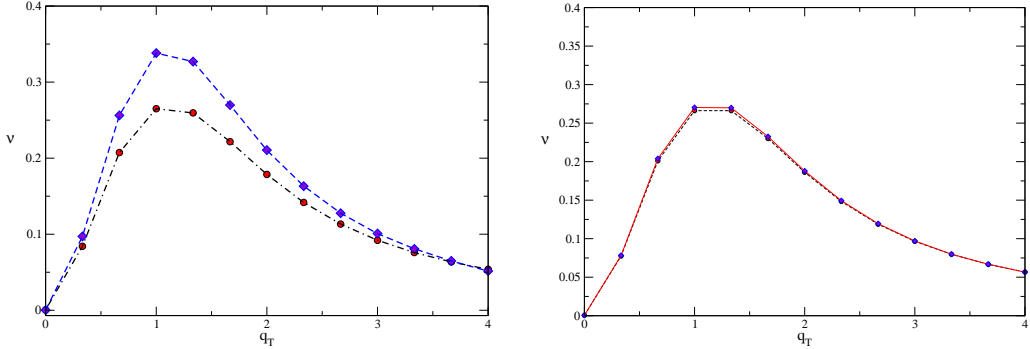

\includegraphics[width=6.5cm]{nu_qt_50.eps}
\hskip 0.5cm
\includegraphics[width=6.5cm]{nu_qt_500.eps}
\caption{Left panel:  $\nu$ plotted as a function of $q_T$ for 
 $s=50\ {\rm GeV}^2$ and $x$ in the range $0.2-1.0$, and  $q$ ranging from
$2.5-5.0 \ {\rm GeV}$: Right panel: 
$s=500\ {\rm GeV}^2$ and $q$ from $4.0-8.6 \ {\rm GeV}$. }
\label{nuqt}
\end{figure}
\vspace{0.5cm}
Collins and Soper considered this effect as a possible source for the 
azimuthal asymmetry that had been measured. It arises from a dependence 
on the relative quark and anti-quark azimuthal orientation that enters in 
the convolution.

We estimate the leading twist $2$ and twist $4$ contributions using
a parton model with the quark diquark spectator framework to 
estimate the $T$-odd and even distribution functions. Previously,
we calculated these functions to predict single spin asymmetries and 
azimuthal dependences 
in SIDIS~\cite{gold_gamb,gamb_gold_ogan1,gamb_gold_ogan2,dis03};
$h_1^\perp$ is given by 
\be
h_1^\perp(x,k_\perp) =
\frac{e_1e_2g^2}{2(2\pi)^4}
\frac{(m+xM)(1-x)}{\Lambda(k^2_\perp)}
\frac{1}{k_\perp^2}
e^{-2b\left(k^2_\perp- \Lambda(0)\right)}
\hspace{-.15cm}\left[\Gamma(0,2b\Lambda(0))\hspace{-.10cm}-\hspace{-.10cm}
\Gamma(0,2b\Lambda(k^2_\perp))\right] , 
\ee
and the unpolarized quark distribution function is
\bea
f(x)&=&\frac{g^2}{(2\pi)^2}\left(1-x\right) 
\cdot \bigg\{\frac{\left(m+xM\right)^2-\Lambda(0)}{\Lambda(0)}
-\left[2b\left(\left(m+xM\right)^2-\Lambda(0)\right)-1\right]
\nn &&\hspace{0.2cm} \times e^{2b\Lambda(0)}\Gamma(0,2b\Lambda(0))\bigg\} .
\eea
We have included a Gaussian damping of the quark transverse momenta in the 
nucleon. This models the known intrinsic $k_\perp$ distribution and also 
regularizes the convolutions that have to be done to obtain observable 
cross sections and asymmetries.
 
\begin{figure}[h]
\includegraphics[height=5.5cm, width=6.5cm]{nu_x_50.eps}
\hskip 0.5cm
\includegraphics[height=5.5cm,width=6.5cm]{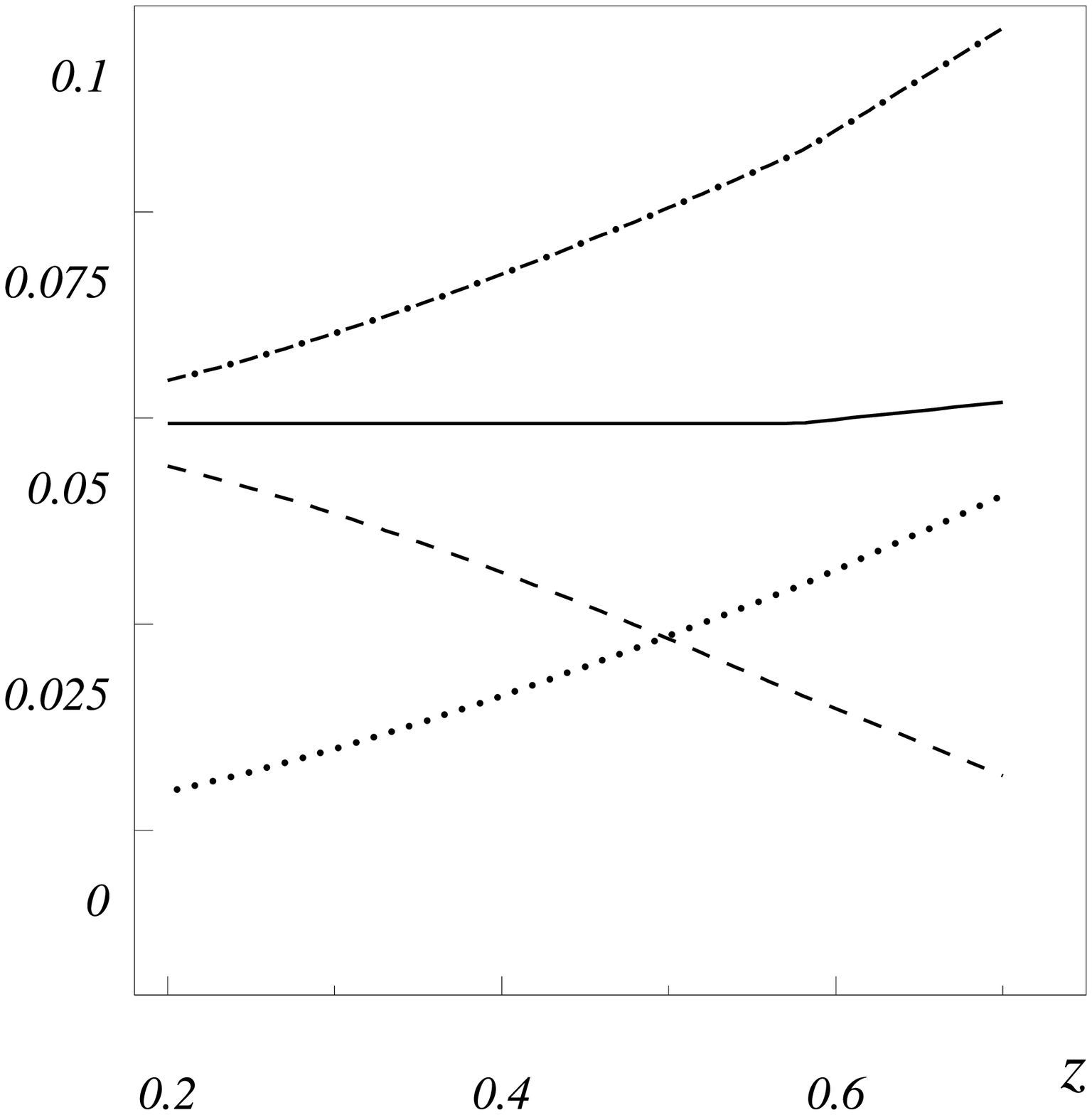}
\caption{Left panel: $\nu$ plotted as a function of $x$ for 
 $s=50\ {\rm GeV}^2$ $q_T$ ranging from 2 to 4 GeV. Right panel:
The $z$-dependence of the $\cos2\phi$ asymmetry at HERMES~\cite{HERMES}
kinematics. The 
full and dotted curves correspond to the $T$-even and $T$-odd terms of
asymmetry, respectively. The dot-dashed and dashed curves are the sum and
the difference of those terms, respectively.}
\label{nux}
\end{figure}

In Fig.~\ref{nuqt},
at center of mass energy of $s=50\ {\rm GeV}^2$,  
the leading order $T$-odd contribution contributes about $28\% $ with an
additional $10\%$ from
sub-leading order $T$-even contributions. 
At center of mass energy of $s=500\ {\rm GeV}^2$ the distinction between
the leading order $T$-odd and additional 
sub-leading order $T$-even contributions
are small. This reflects the diminution of the non-leading contribution 
with increasing s and $Q^2$. Note that we have not taken account of the
 evolution of the distribution $h_1^{\perp}$ with $Q^2$ scale. This 
evolution has not been worked out in general at this time. 
In  Fig.~\ref{nux}, $\nu$ is plotted versus $x$ at $s=50\ {\rm GeV}^2$ 
where $q_T$ ranges from 2 to 4 GeV.  Again the higher twist contribution is
significant.
\vspace{0.7cm}
\vspace{0.5cm}
\subsubsection{$\cos 2\phi$ Azimuthal Asymmetry in SIDIS}
We use the conventions established in~\cite{BM} for the
asymmetries.  Being $T$-odd, $h_1^\perp$ appears with the $H_1^\perp$, 
the $T$-odd fragmentation function in observable quantities.
In particular, the following weighted 
SIDIS cross section projects out a leading double $T$-odd
$\cos2\phi$ asymmetry, 
\bea
{\langle  \cos2\phi \rangle}
{\scriptscriptstyle_{UU}}&=& 
\frac{{8(1-y)} \sum_q e^2_q h^{\perp(1)}_1(x) z^2 H^{\perp(1)}_1(z)}
{{(1+{(1-y)}^2)}  \sum_q e^2_q f_1(x) D_1(z)}
\label{ASY}, 
\eea
where the subscript $UU$ indicates unpolarized beam and 
target. The Collins fragmentation function is given by~\cite{gamb_gold_ogan2},
\bea
H_1^\perp(z,k_\perp)=\hspace{-.15cm} 
\frac{f_{qq\pi}^2g^2}{(2\pi)^4}
\frac{M\mu/z}{\Lambda(k^2_\perp)}\frac{(1-z)}{4z^2}
\frac{1}{k_\perp^2}
e^{-2c\left(k^2_\perp- \Lambda(0)\right)}
\hspace{-.15cm}\left[\Gamma(0,2c\Lambda(0))\hspace{-.10cm}-\hspace{-.10cm}
\Gamma(0,2c\Lambda(k^2_\perp))\right] .
\eea
The   $\cos2\phi$ 
asymmetry originating from $T$-even distribution and 
fragmentation function 
appears at order $1/Q^2$~\cite{CAHN,LM}. 
The $\langle \cos2\phi \rangle$ from ordinary sub-sub-leading 
$T$-even and leading double $T$-odd (up to a sign) effects 
to order $1/Q^2$ is given by~\cite{dis03}
\begin{equation}
{\langle \cos2\phi \rangle}_{UU} 
=\frac{2\frac{\langle k^2_{\perp }\rangle}{Q^2} (1-y) f_1(x)D_1(z) 
\pm 8 (1-y) h^{\perp (1)}_1(x) H^{\perp (1)}_1(z)}{ \bigg [ 1+{(1-y)}^2 + 
2\frac{\langle k^2_{\perp }\rangle}{Q^2} (1-y) \bigg ] f_1(x) D_1(z)}.
\label{C2PHI}
\end{equation}   
The $z$-dependence of this asymmetry at HERMES kinematics~\cite{HERMES} 
are shown in the right panel of Fig.~\ref{nux}. The 
full and dotted curves correspond to the $T$-even and $T$-odd terms in the
asymmetry, respectively. The dot-dashed and dashed curves are the sum and
the difference of those terms, respectively. From the figure, one can 
see that the double 
$T$-odd asymmetry behaves like $z^2$, while the $T$-even 
asymmetry is flat in the whole range of $z$.  
Therefore, aside from the
competing $T$-even $\cos2\phi$ effect,  the experimental 
observation of a strong $z$-dependence (especially at high $z$ region) 
would indicate the presence of $T$-odd structures in {\it unpolarized}  
SIDIS implying that novel transversity properties of the  nucleon 
can be accessed  without involving spin polarization.     

\subsubsection{Conclusions}
The interdependence of intrinsic transverse
quark momentum and angular momentum conservation are intimately
connected with studies of $T$-odd effects underlying the $\cos 2\phi$
azimuthal asymmetries in Drell-Yan and semi-inclusive deep inelastic
scattering. 
Using re-scattering as a mechanism to generate 
$T$-odd distribution functions 
opens a new window into the theory and phenomenology of transversity in 
hard processes.  We have also demonstrated that at moderate energies sub-leading twist contributions are non-trivial.


\subsubsection{Acknowledgments}
L.P.G. thanks J. Soffer, C. Bourrely and Z.E. Meziani
and the organizers of the  HiX2004 Workshop 
for the opportunity to present this work. 







\end{document}